\documentclass{PoS}
\newcommand{\hsp}{\hspace*{1pt}}
\title{Hydrodynamical modeling of the deconfinement phase transition and 
explosive hadronization}

\ShortTitle{Hydrodynamical modeling}

\author{\speaker{Igor N. Mishustin}\thanks{A footnote may follow.}\\
        Frankfurt Institute for Advanced Studies, J.W. Goethe University,
D--60438 Frankfurt am Main, Germany\\
Russian Research Center ''Kurchatov Institute'', 123182 Moscow, Russia\\
 E-mail: \email{mishustin@fias.uni-frankfurt.de}}

\abstract{Dynamics of relativistic heavy-ion collisions is investigated on the basis of a 
simple (1+1)-dimensional hydrodynamical model in light-cone coordinates. The 
main emphasis is put on studying sensitivity of the dynamics and observables to 
the equation of state and initial conditions. Low sensitivity of pion rapidity spectra to the presence of the phase transition is demonstrated, and some inconsistencies of the equilibrium scenario are pointed out. Possible non-equilibrium effects are discussed, in particular, a possibility of an explosive disintegration of the deconfined phase into quark-gluon droplets. Simple estimates show that the characteristic droplet size should decrease with increasing the collective expansion rate. These droplets will hadronize individually by emitting hadrons from the surface. This scenario should reveal itself by strong non-statistical fluctuations of observables.}

\FullConference{Critical Point and Onset of Deconfinement
          4th International Workshop\\
		 July 9-13 2007\\
		 GSI Darmstadt,Germany}

\begin{document}

\section{Introduction}
High--energy heavy--ion collisions provide a unique tool for studying
properties of hot and dense strongly--interacting matter in the
laboratory. The theoretical description of such collisions is often
done within the framework of a hydrodynamic approach.  This approach
opens the possibility to study the sensitivity of collision dynamics
and secondary particle distributions to the equation of state (EOS) of
the produced matter. The two most famous realizations of this approach,
which differ by the initial conditions, have been proposed by
Landau~\cite{Lan53} (full stopping) and Bjorken~\cite{Bjo83} (partial
transparency). In recent decades many versions of the hydrodynamic
model were developed.

Below we apply a simplified version of the hydrodynamical
model, dealing only with the longitudinal dynamics of the fluid (see 
details in refs.~\cite{Sat06,Sat07}). This
approach has as its limiting cases the Landau and Bjorken models. We
investigate the sensitivity of the hadron rapidity spectra to the
fluid's equation of state, to the choice of initial state and
freeze--out conditions. Special attention is
paid to possible manifestations of the deconfinement phase transition.
In particular, we compare the dynamical evolution of the fluid and 
secondary particle spectra calculated with and without the phase transition.
In the second part of the talk I present arguments in favour of the explosive 
hadronization of the quark-gluon plasma, first formulated in ref.~\cite{Mis99}. 

\section{Hydrodynamical equations in light-cone coordinates}

We consider central collisions of equal nuclei disregarding the effects
of transverse collective expansion. In this case one can parameterize the 
collective fluid velocity, $U^\mu$, in terms of the longitudinal flow 
rapidity~$Y$ as $U^\mu=(\cosh{Y},0,0,\sinh{Y})^\mu$. It is convenient to 
make transition from the usual space--time coordinates $t,z$ to
the hyperbolic (light--cone) variables, namely,
the proper time~$\tau$ and the space--time rapidity $\eta$\,, defined as
\begin{equation} \label{lcva}
\tau=\sqrt{t^2-z^2}\,,\hspace*{5mm}\eta=\tanh^{-1}\left(\frac{z}{t}\right)=
\frac{1}{2}\ln{\frac{t+z}{t-z}}\,.
\end{equation}
In these coordinates the hydrodynamic equations take the
following form
\begin{eqnarray}
\left[\tau\frac{\partial}{\partial\tau}+
\tanh\hsp (Y-\eta)\hsp\frac{\partial}{\partial\eta}\right]\hsp n
+n\left[\tanh\hsp (Y-\eta)\hsp\tau\frac{\partial}{\partial\tau}+
\frac{\partial}{\partial\eta}\right] Y&=&0\,,
\label{bcce1}\\
\left[\tau\frac{\partial}{\partial\tau}+
\tanh\hsp (Y-\eta)\hsp\frac{\partial}{\partial\eta}\right]\hsp\epsilon
+(\epsilon +P)\left[\tanh\hsp (Y-\eta)\hsp\tau\frac{\partial}{\partial\tau}+
\frac{\partial}{\partial\eta}\right] Y&=&0\,,
\label{emte1}\\
(\epsilon +P)\left[\tau\frac{\partial}{\partial\tau}+\tanh\hsp
(Y-\eta)\hsp\frac{\partial}{\partial\eta}\right] Y
+\left[\tanh\hsp (Y-\eta)\hsp\tau\frac{\partial}{\partial\tau}+
\frac{\partial}{\partial\eta}\right] P&=&0\,,
\label{emte2}
\end{eqnarray}
where ${\rm n}$, $\epsilon$ and $P$ are the baryon density, energy 
density and pressure of the fluid.
To solve Eqs.~(\ref{bcce1})--(\ref{emte2}), one needs to specify the
EOS, \mbox{$P=P\hsp (n,\epsilon)$}, and the initial profiles
$n\hsp (\tau_0,\eta)$, $\epsilon\hsp (\tau_0,\eta)$, $Y(\tau_0,\eta)$
at a time $\tau=\tau_0$\, when the fluid may be considered as
thermodynamically equilibrated.

\section{Equation of state}

In this paper we consider only the baryon--free matter, i.e. assume
vanishing net baryon density $n$ and chemical potential $\mu$. In this
case Eq.~(\ref{bcce1}) is trivially satisfied and all thermodynamic
quantities, e.g. pressure, temperature $T$ and entropy density
$s=(\epsilon+P-n\mu)/T$\,, can be regarded as functions of $\epsilon$
only.

\begin{figure*}[htb!]
\centerline{\includegraphics[width=0.8\textwidth]{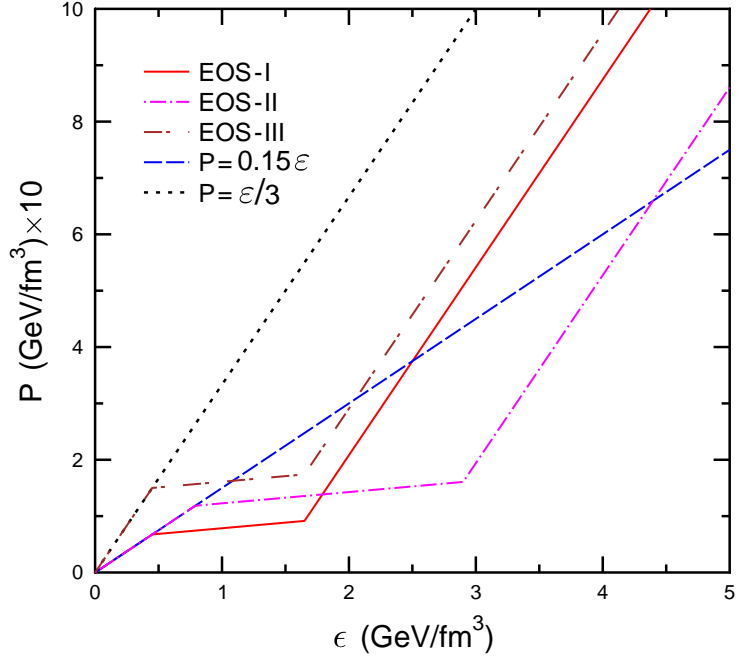}}
\caption{
Comparison of different EOSs used in hydro simulations. Parameters characterizing  EOS--I,
EOS--II and EOS--III are given in ref.~\cite{Sat07}.
}
\label{fig:eps-pres}
\end{figure*}

One of the main goals of experiments on ultrarelativistic heavy--ion
collisions is to study the deconfinement phase transition of
strongly--interacting matter. In our calculations this phase transition is
implemented through a bag--like
EOS using the parameterization suggested in Ref.~\cite{Tea01}. This EOS
consists of three parts, denoted below by indices $H, M, Q$\,, and
corresponding, respectively, to the hadronic, ''mixed'' and
quark--gluon phases. As already mentioned, pressure, temperature and
sound velocity, $c_s=\sqrt{dP/d\epsilon}$, of the baryon--free matter
can be regarded as functions of $\epsilon$\, only. It is further assumed
that $c_s$ is constant in each phase and, therefore, $P$ is a linear function of
$\epsilon$ with different slopes in the corresponding regions of
energy density.

The hadronic phase corresponds to the domain of low
energy densities, \mbox{$\epsilon<\epsilon_H$}, and temperatures,
$T<T_H$\hsp . This phase consists of pions, kaons, baryon--antibaryon pairs
and hadronic resonances. Numerical calculations for the ideal gas of
hadrons (see e.g.~\cite{Cho05}) predict a rather soft EOS:  the corresponding
sound velocity squared, $c_s^2=c_H^2\sim 0.1-0.2$\,, is noticeably lower than 1/3.
The mixed phase takes place at intermediate energy densities, from
$\epsilon_H$ to~$\epsilon_Q$ or at temperatures from~$T_H$
to~$T_Q$\,. The quantity $\epsilon_Q-\epsilon_H$ can be interpreted as
the ''latent heat'' of the deconfinement transition. To avoid numerical problems, we
choose a small, but nonzero value of sound velocity $c_M$ in the mixed
phase. The third, quark--gluon plasma region of the EOS corresponds to
$\epsilon>\epsilon_Q$ or $T>T_Q$. It is assumed that $c_s^2=c_Q^2$ reaches
the asymptotic value (1/3) already at the beginning of the quark--gluon
phase, i.e. at $T\simeq T_Q$\,.

\begin{figure*}[htb!]
\centerline{\includegraphics[width=0.8\textwidth]{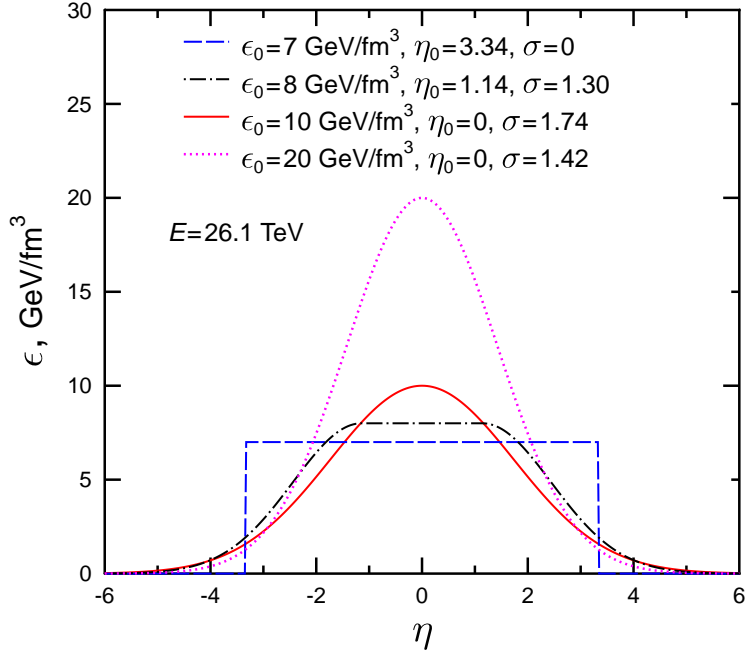}}
\caption{
Initial energy density profiles used in the hydro simulations. Solid and 
dashed--dotted lines represent, respectively, the parameter sets A and~C. 
Dashed line is calculated for the table--like profile with $\eta_0=3.34$\,, 
$\epsilon_0=7$\,GeV/fm$^3$. Dotted line corresponds to the Gaussian
profile with $\sigma=1.42$\,, $\epsilon_0=20$\,GeV/fm$^3$\,.
}
\label{fig:eta-epsi}
\end{figure*}

Figure \ref{fig:eta-epsi} represents some profiles of initial energy
density used in our calculations. They are parameterized by a flattened Gaussian 
distributions with a plateau at $-\eta_0\leq \eta \leq \eta_0$ and wings of width 
$\sigma$. All these profiles correspond to the same
total energy of secondaries, $E\approx 26$ Tev, as estimated by the BRAHMS 
collaboration for Au+Au collisions at RHIC energy 
$\sqrt{s_{\scriptscriptstyle NN}}=200$\,GeV \cite{BRAHMS}. One can see that
three different phases of matter appear already at the initial state.

\section{Hydrodynamical evolution of matter}

\begin{figure*}[htb!]
\hspace*{-15mm}\includegraphics[width=0.9\textwidth]{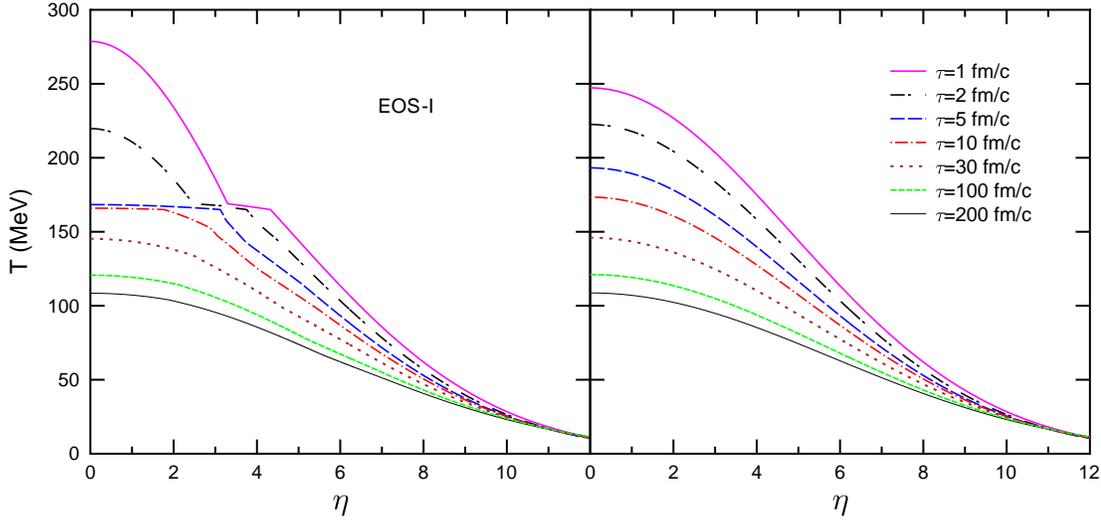}
\caption{
Temperature profiles at different proper times $\tau$ calculated for
initial conditions corresponding to the parameter set A.
Only the forward hemisphere ($\eta\geqslant 0$) is shown. Left and
right panels correspond, respectively, to the EOS--I and the pure hadronic
EOS $P=c_H^2\hsp\epsilon$\, with $c_H^2=0.15$\,.
}
\label{fig:eta-tem1}
\end{figure*}
Space--time evolution of matter as predicted by the present model is
illustrated in Fig.~\ref{fig:eta-tem1}.
This figure shows profiles of the
temperature at different proper times
$\tau$\,. Here we consider the Gaussian--like initial conditions, with
parameters from the set A (see  Fig.~\ref{fig:eta-epsi}. For comparison, 
the results  are presented for the
EOS--I and for the hadronic EOS with $c_s^2=0.15$. One can see
that in the case of the phase transition the model predicts appearance
of a a flat shoulder in $T\hsp(\eta)$
which is clearly visible at $\tau\lesssim 10$\,fm/c. This is a
manifestation of the mixed phase which exists during the time interval
$\Delta\tau\sim 10$\,fm/c. According to Fig.~\ref{fig:eta-tem1}, the largest
volume of this phase in the $\eta$--space is formed at $\tau\sim 5$\,fm/c.
In the considered case the ''memory'' of the quark-gluon  phase is practically
washed out at $\tau\gtrsim 30$\,fm/c. 

\begin{figure*}[htb!]
\hspace*{-15mm}\includegraphics[width=0.9\textwidth]{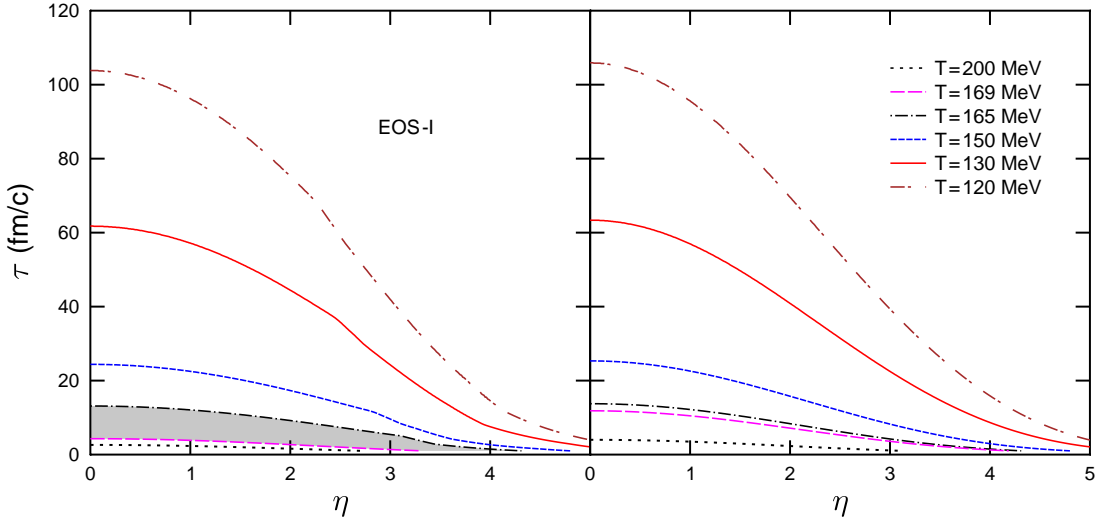}
\caption{
Isotherms in the $\eta-\tau$ plane calculated for the parameter set A.
Left and right panels correspond to the same EOS\hsp s and initial conditions 
as in Fig.~3. Shaded region indicates the mixed phase.
}
\label{fig:eta-tauf}
\end{figure*}


According to Fig.~\ref{fig:eta-tauf}
(see the left panel), the initial stage of the evolution when matter is in the
quark--gluon phase, lasts only for a very short time, of about 5 fm/c.
The region of the mixed phase is crossed in less than 10 fm/c. This
clearly shows that the slowing down of expansion associated with the
''soft point'' of the EOS \cite{Shu,Ris} plays no role, when the initial state lies
much higher in the energy density than the phase transition region. In
this situation the system spends the longest time in the hadronic
phase. 

\begin{figure*}[htb!]
\hspace*{-15mm}
\includegraphics[width=0.9\textwidth]{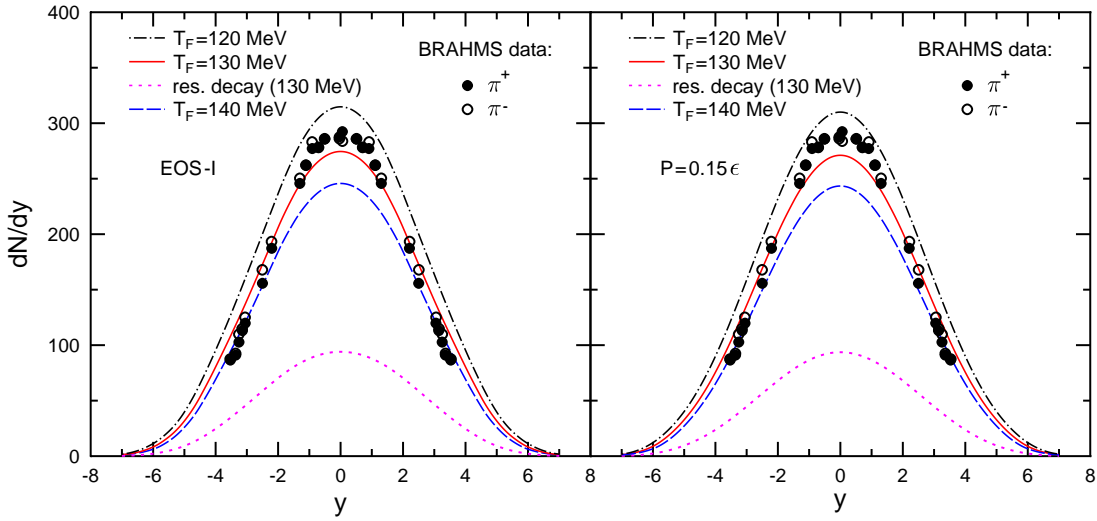}
\caption{
Rapidity distribution of $\pi^+$--mesons in central Au+Au collisions at
$\sqrt{s_{\scriptscriptstyle NN}}=200$\,GeV. Left panel shows
results of hydrodynamical
calculations for the EOS--I and initial conditions with parameters
of set A. Right panel corresponds to the pure hadronic EOS with $c_H^2$=0.15 
and the same initial conditions. Solid, dashed and dashed--dotted curves correspond to
different values of the freeze--out temperature $T_F$\,. The dotted
line shows contribution of resonance decays in the case $T_F=130$\,MeV.
Experimental data are taken from Ref.~\cite{Bea05}.
}
\label{fig:rap-pis1}
\end{figure*}

This low sensitivity to the EOS is clearly seen in Fig.~\ref{fig:rap-pis1}, 
which shows the rapidity distributions of secondary pions. One can see that two
EOS with and without the phase transition lead to very similar observable 
pion rapidity spectra, which both agree very well with the BRAHMS data \cite{Bea05}. The
best fit of experimental data is achieved with freeze-out temperature $T_F=130$\,MeV. 
The feeding from resonance decays was accurately taken into account (see details in ref.~\cite{Sat07}).

As one can see in Fig.~\ref{fig:eta-tauf}, the freeze-out at $T_F=130$\,MeV requires 
an expansion time of about 60 fm/c at $\eta=0$. This is certainly a very long time which is
seemingly in contradiction with some experimental findings. Indeed, the
interferometric measurements~\cite{Adl01} show much shorter times of
hadron emission, of the order of 10 fm/c.
As follows from our results, this discrepancy can not be
removed by considering other EOS or initial conditions. A considerable
reduction of the freeze--out times can be achieved by including the
effects of transverse expansion and chemical
nonequilibrium~\cite{Hir02}. However, this will not change essentially
the dynamics of the early stage ($\tau\lesssim 10$\,fm/c) when
expansion is predominantly one--dimensional. A more radical solution
would be an explosive decomposition of the quark--gluon plasma,
proposed in Ref.~\cite{Mis99}. This may happen at very early times,
right after crossing the critical temperature line, when the plasma
pressure becomes very small or negative. We shall consider this
possibility in the second part of the talk.

\section{Explosive hadronization}

Let us consider now a simplified picture where the system expands according to 
the Hubble law, $v(r)=H\cdot r$, where $v$ is the local collective velocity and $H$ 
is a function of time, as e.g. $H\propto 1/t$, in the Bjorken model.

As demonstrated in refs.~\cite{Mis99,Mis05}, a first order phase transition in 
rapidly expanding system will not follow the phase equilibrium trajectory. Instead, the 
high-temperature phase will expand until it enters the spinodal region. Then, due to 
intrinsic instabilities it will disintegrate into droplets surrounded by the 
undersaturated low-temperature phase. Different aspects of spinodal decomposition 
in expanding systems were discussed in refs. \cite{Cse95, Sca01,Ran04}.
For clarity, below we use capital letters 
Q and H (not to be confused with the Hubble constant $H$) for the 
deconfined quark-gluon phase and the hadronic phase, respectively.
Following this picture, let us assume that the 
dynamical fragmentation of the deconfined phase has resulted in a collection of Q droplets
embedded in a dilute H phase, as illustrated in Fig.~6. 
The optimal droplet size can be determined by applying a simple energy balance
argument saying that the droplets are formed when the collective kinetic energy within the
individual droplet is equal to its surface energy, $E_{\rm kin}(R)=E_{\rm surf}(R)$, where 
\begin{equation}
E_{\rm kin}(R)=\frac{1}{2}\int_0^R\Delta{\cal E}[v(r)]^24\pi r^2dr=
\frac{2\pi}{5}\Delta{\cal E}H^2R^5,
\end{equation}
and $E_{\rm surf}(R)=4\pi R^2\gamma$, where $\Delta{\cal E}={\cal E}_{\rm Q}-{\cal E}_{\rm H}$ 
is the energy density difference of the two phases, and $\gamma$ is the corresponding surface tension. Then the optimal droplet size is given by the expression    
\begin{equation}  \label{R}
R^*=\left(\frac{10\gamma}{\Delta{\cal E}H^2}\right)^{1/3}.
\end{equation}

\begin{figure*}[htp!]
\includegraphics[width=0.9\textwidth]{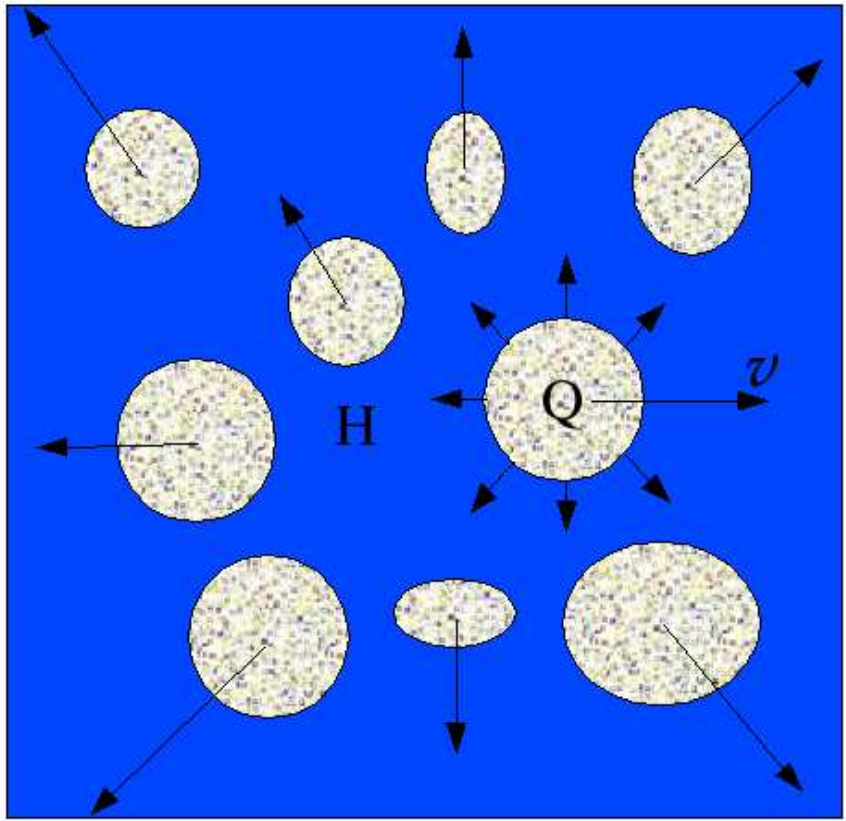}
\caption{(Color online) Schematic view of multi-droplet state 
produced after the dynamical
fragmentation of a metastable high energy-density Q phase. 
The Q droplets are embedded in the low energy-density H phase. 
Each droplet emits hadrons as a thermal source, as well as 
participates in the overall Hubble-like expansion.} 
\label{potcrit}
\end{figure*}

\noindent As Eq. (\ref{R}) indicates, the droplet size depends 
strongly on $H$. When expansion is slow (small $H$) the droplets are big.
In the adiabatic limit the process may look like a fission of a cloud of plasma.
But fast expansion should lead to very small droplets. This state of matter 
is very far from a thermodynamically
equilibrated mixed phase, particularly because the H phase 
is very dilute. One can say that the metastable Q matter is torn apart by a 
mechanical strain associated with the collective expansion. 

The driving force for expansion is the pressure gradient, 
$\nabla P
\equiv c_s^2\nabla {\cal E}$,
which depends crucially on the sound velocity in matter, $c_s$.
Here we are interested in the expansion rate of the partonic phase, which 
is not directly observable but predicted by the hydrodynamical simulations.
In the vicinity of the phase transition, one should expect a ``soft point'' 
\cite{Shu,Ris} where the sound velocity is smallest and the ability of 
matter to generate the collective expansion is minimal. If the initial state 
of the Q phase is close to this point, its subsequent expansion will be slow.
Accordingly, the droplets produced in this case will be big. When moving away 
from the soft point, one would see smaller and smaller droplets. For numerical 
estimates we choose two values of the Hubble constant: $H^{-1}$=20 fm/c to 
represent the slow expansion from the soft point and $H^{-1}$=6 fm/c for the 
fast expansion.  

One should also specify two other parameters, $\gamma$ and 
$\Delta{\cal E}$. The surface tension $\sigma$ is a subject of debate at 
present. Lattice simulations indicate that it could be as low as a few MeV/fm$^2$
in the vicinity of the critical line. However, for our non-equilibrium scenario, 
more appropriate values are closer to 10-20\,MeV/fm$^2$, which follow from 
effective chiral models. As a compromise, the value $\gamma=10$\,MeV/fm$^2$ is 
used below for rough estimates. Bearing in mind that nucleons and heavy mesons 
are the smallest 
droplets of the Q phase, one can take $\Delta{\cal E}=0.5$\,GeV/fm$^3$, i.e. 
the energy density inside the nucleon. Then one gets $R^*=3.4$\,fm for 
$H^{-1}=20$\,fm/c and $R^*=1.5$\,fm for $H^{-1}=6$\,fm/c. 
As follows from eq.~(\ref{R}), for a spherical droplet $V\propto 1/\Delta{\cal E}$,
and in the first approximation its mass, 
\begin{equation}
M^*\approx \Delta{\cal E}V=\frac{40\pi}{3}\frac{\gamma}{H^2},
\end{equation} 
is independent of $\Delta{\cal E}$. For the two values of $R^*$ given above the optimal 
droplet mass is $\sim 100$\,GeV and $\sim 10$\,GeV, respectively. 
As shown in ref. \cite{Mis99}, the 
distribution of droplet masses should follow an exponential law, $\exp{\left(-{M 
\over M^*}\right)}$. Thus, about 2/3 of droplets have masses smaller than
$M^*$, but with 1$\%$ probability one can find droplets as heavy as $5M^*$. 

In refs. \cite{Mis1,Dum03} the evolution of individual droplets was 
studied numerically within a 
hydrodynamical approach including dynamical chiral fields (Chiral Fluid Dynamics).  
It has been demonstrated that the energy released at the spinodal decomposition 
can be transferred directly into the collective oscillations of the ($\sigma,{\bf \pi}$) 
fields which give rise to the soft pion radiation. However,
eventually the Q droplets will hadronize by emitting hadrons from the surface.
This scenario can explain short emission times of hadrons observed in experiments, 
see e.g. ref. \cite{Adl01}.

\section{Anomalous multiplicity fluctuations}

\begin{figure*}[htp!]
\includegraphics[width=0.9\textwidth]{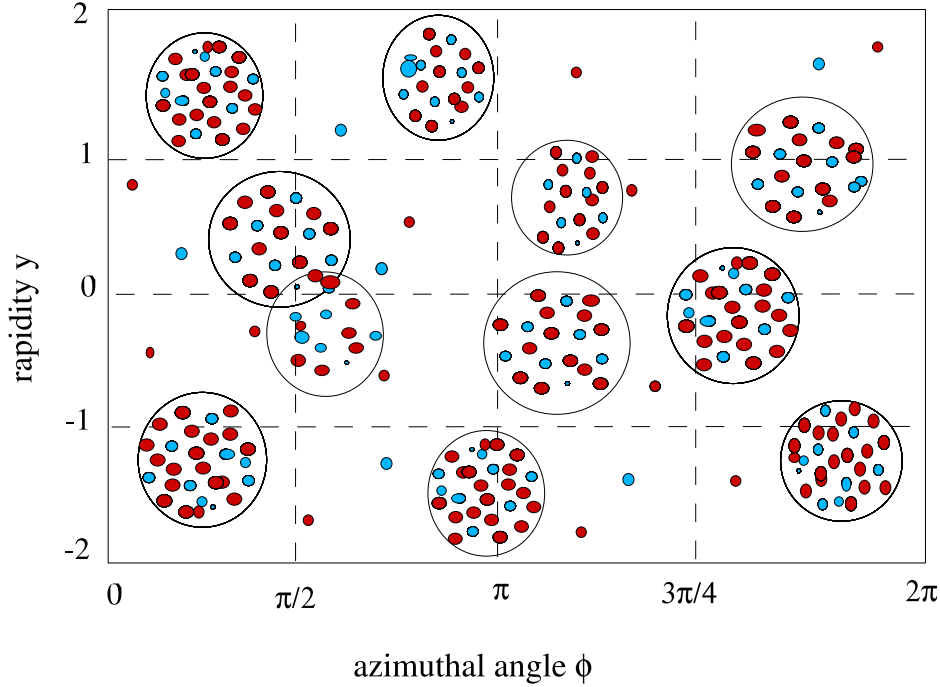}
\caption{(Color online) Schematic view of the momentum space distribution of secondary 
hadrons produced from an ensemble of droplets. Each droplet emits hadrons 
(mostly pions) within a rapidity interval $\delta y\sim1$ and azimuthal 
angle spreading of $\delta\phi\sim 1$.} 
\label{potcrit}
\end{figure*}

After separation the QGP droplets recede from each other according to the
global expansion, pre\-do\-mi\-nant\-ly along the beam direction.
Hence their center-of-mass rapidities $y_i$ are in one-to-one correspondence
with their spatial positions.
Presumably $y_i$ will be distributed more or less evenly between
the target and projectile rapidities. 
Since rescatterings in the dilute H phase between the droplets are rare, most hadrons 
produced from individual droplets will go directly into detectors. This may explain why 
freeze-out parameters extracted from the hadronic yields are always very close 
to the phase transition boundary \cite {And06}.    

In the droplet phase the mean number of produced hadrons in a given rapidity 
interval is
$\langle N\rangle=\sum\limits_i^{N_D}\overline{n_i}=\langle n\rangle\langle N_D\rangle$,
where $\overline{n_i}$ is the mean multiplicity of hadrons emitted from a 
droplet i, $\langle n\rangle$ is the average multiplicity per droplet and 
$\langle N_D\rangle $ is the mean number of droplets produced in this interval.
If droplets do not overlap in the rapidity space, each
droplet will give a bump in the hadron rapidity distribution around its
center-of-mass rapidity $y_i$ \cite{Cse95,Mis99,Mis05}. In case of the Boltzmann spectrum the
width of the bump will be $\delta \eta \sim \sqrt{T/m}$, where $T$ is the
droplet temperature and $m$ is the particle mass. At $T\sim 100$ MeV this
gives $\delta \eta \approx 0.8$ for pions and $\delta \eta \approx 0.3$ for nucleons.
These spectra might be slightly modified by the residual expansion of droplets.
Due to the radial expansion of the fireball the droplets should also be well 
separated in the azimuthal angle. The characteristic angular spreading of pions
produced by an individual droplet is determined by the ratio of the thermal 
momentum of emitted pions to their mean transverse momentum, $\delta\phi\approx 
3T/\langle p_{\perp}\rangle\sim$ 1. 
The resulting phase-space distribution of hadrons in a single event will be a 
superposition of contributions from different Q droplets superimposed on a more
or less uniform background from the H phase. Such a distribution is shown schematically 
in Fig.~7. It is obvious that such inhomogeneities (clusterization) in the 
momentum space will reveal strong non-statistical fluctuations.
The fluctuations will be more pronounced if primordial droplets are big,
as expected in the vicinity of the soft point. If droplets as heavy as 100 GeV are 
formed, each of them will emit up to $\sim$200 pions within a narrow rapidity and 
angular intervals, $\delta \eta \sim 1$, $\delta\phi\sim 1$. If only a few droplets 
are produced in average per unit rapidity, $N_D\gtrsim 1$, 
they will be easily resolved and analyzed. On the other hand, the fluctuations 
will be suppressed by factor $\sqrt{N_D}$ if many small droplets shine into the same 
rapidity interval.

It is convenient to characterize the fluctuations by the scaled variance
$\omega_N\equiv (\langle N^2\rangle-\langle N\rangle^2)/\langle N\rangle$.
Its important property is that $\omega_N=1$ for the Poisson distribution, and therefore 
any deviation from unity will signal a non-statistical emission mechanism.
As shown in ref. \cite{Bay}, for an ensemble of emitting sources (droplets) 
$\omega_N$ can be expressed in a simple form, 
$\omega_N=\omega_n+\langle n\rangle\omega_D$, 
where $\omega_n$ is an average multiplicity fluctuation in a single droplet,
$\omega_D$ is the fluctuation in the droplet size distribution and $\langle n\rangle$ 
is the mean multiplicity from a single droplet. Since $\omega_n$ and $\omega_D$ 
are typically of order of unity, the fluctuations from the multi-droplet emission
are enhanced by the factor $\langle n\rangle$. According to the picture of
a first order phase transition advocated above, this enhancement factor could
be as large as $\sim 10$. It is clear that the nontrivial structure of the hadronic 
spectra will be washed out to a great extent when averaging over many events. 
Therefore, more sophisticated methods of the event sample analysis should be 
applied as e.g. measuring event-by-event fluctuations in the hadron multiplicity 
distributions in a varied rapidity bin. Up to now no significant effects in fluctuation observables have been found \cite{Kon07}. 

\section{Conclusions}

\begin{itemize}

\item Rapidity distributions of pions and kaons at RHIC energy can be well described by the ideal hydro with a soft EOS and initial energy density of 5-10 GeV/fm$^3$.

\item Equilibrium hydrodynamics is not sensitive to a phase transition if the initial state is far from the transition point. In this case two EOS$\hsp$s with and without the phase transition 
give similar results for observables.  

\item To explain short emission times observed in experiments one may assume an explosive disintegration of the quark-gluon plasma at the phase transition boundary.
This should lead to the formation of quark-gluon droplets which will manifest themselves in non-statistical fluctuations of observables.

\item Better conditions for observation of the deconfinement phase transition may occur at lower energies when the baryon density is higher but the initial pressure is lower. The future FAIR facility at GSI should be a right place to search for manifestations of the deconfinement phase transition in baryon-rich environment.  

\end{itemize}

The author thanks L.M. Satarov and H. St\"ocker for the fruitful collaboration on the hydrodynamic modeling of relativistic heavy-ion collisions. I am also grateful to Igor 
Pshenichnov for the help in preparation of this talk. This work was supported in part by
the BMBF, GSI, the DFG grant~~436 RUS ~~~\mbox{113/711/0--2} (Germany),
and the grants RFBR 05--02--04013 and NS--8756.2006.2 (Russia).

\end{document}